\documentclass[a4paper,fleqn]{cas-dc}

\usepackage[numbers]{natbib}

\usepackage{eurosym,makecell,comment}
\usepackage{multirow}
\usepackage{threeparttable}
\usepackage{framed}
\usepackage{hyperref}
\usepackage{url}
\usepackage{verbatim}
\usepackage{graphicx}
\usepackage{subcaption}
\usepackage{float}
\usepackage{xcolor}
\usepackage{microtype}
\usepackage{colortbl}
\usepackage{amsmath,amssymb,amsfonts}
\usepackage{tikz}
\usepackage{pgfplots}
\usepackage{listings}

\usepackage{pifont}
\usepackage{algorithm}
\usepackage{algpseudocode}
\usepackage{balance}

\pgfplotsset{compat=1.18} 

\newcommand{\xmark}{\ding{55}}%

\newcommand{\ie}{\textit{i.e., }}  
\newcommand{\eg}{\textit{e.g., }}  

\def \1{\textit{(i)}}
\def \2{\textit{(ii)}}
\def \3{\textit{(iii)}}
\def \4{\textit{(iv)}}
\def \5{\textit{(v)}}

\usepackage{etoolbox}
\usepackage{soul} 

\newtoggle{finalPaper}

\setstcolor{blue}

\toggletrue{finalPaper} 

\iftoggle{finalPaper} {

	\newcommand{\rmvtxt}[1]{}}
{

	\newcommand{\rmvtxt}[1]{\st{#1}}}

\newtoggle{removeItalic}
\togglefalse{removeItalic} 
\iftoggle{removeItalic} {
    \renewcommand{\textit}[1]{#1}
}

\newcommand{\solution}{\textit{GuardFS}}

\begin{document}
\let\WriteBookmarks\relax
\def\floatpagepagefraction{1}
\def\textpagefraction{.001}
\shorttitle{Integrated Detection and Mitigation of Ransomware Through the File System}
\shortauthors{von der Assen et~al.}

\title[mode = title]{\textit{GuardFS}: a File System for Integrated Detection and Mitigation of Linux-based Ransomware}

\author[1]{Jan {von der Assen}}[orcid=0000-0002-0591-8887]
\author[1]{Chao Feng}[orcid=0000-0002-0672-1090]
\author[1]{Alberto {Huertas Celdrán}}[orcid=0000-0001-7125-1710]
\author[1]{Róbert Oleš} []
\author[2]{Gérôme Bovet} [orcid=0000-0002-4534-3483]
\author[1]{Burkhard Stiller}[orcid=0000-0002-7461-7463]

\address[1]{Communication Systems Group CSG, Department of Informatics IfI, University of Zurich UZH, CH---8050 Zürich, Switzerland}

\address[2]{Cyber-Defence Campus within armasuisse Science \& Technology, CH---3602 Thun, Switzerland}

\cortext[cor1]{Corresponding author.
Email address: vonderassen@ifi.uzh.ch (J. von der Assen)}

\begin{keywords}
Cybersecurity \sep Ransomware \sep Malware \sep Fingerprinting \sep Machine Learning

\end{keywords}

\maketitle

\begin{abstract}
Although ransomware has received broad attention in media and research, this evolving threat vector still poses a systematic threat. Related literature has explored their detection using various approaches leveraging Machine and Deep Learning. While these approaches are effective in detecting malware, they do not answer how to use this intelligence to protect against threats, raising concerns about their applicability in a hostile environment. Solutions that focus on mitigation rarely explore how to prevent and not just alert or halt its execution, especially when considering Linux-based samples. This paper presents \textit{GuardFS}, a file system-based approach to investigate the integration of detection and mitigation of ransomware. Using a bespoke overlay file system, data is extracted before files are accessed. Models trained on this data are used by three novel defense configurations that obfuscate, delay, or track access to the file system. The experiments on \textit{GuardFS} test the configurations in a reactive setting. The results demonstrate that although data loss cannot be completely prevented, it can be significantly reduced. Usability and performance analysis demonstrate that the defense effectiveness of the configurations relates to their impact on resource consumption and usability.

\end{abstract}

\section{Introduction}
\label{sec:intro}
Digitalization has become pervasive in today's business landscape. The technological advancements that uphold today's enterprises have brought many advantages, such as reduced operational costs, improved time-to-market strategies, and global access to suppliers and customers. However, this also made organizations vulnerable to cybersecurity breaches. As past instances have shown, the impact of successful breaches can take many forms, ranging from economic damage to the endangerment of personal health~\cite{cisa, coretm}. 

Within the lively threat landscape, many attack vectors are employed. One threat that has been evolving over multiple decades of proliferation is ransomware. Ransomware distributed over floppy drives has been mentioned as early as the 1980s~\cite{ransomware-history}. Nevertheless, ransomware was not a widely discussed attack until the emergence of digital payment solutions in the 2010s. Nowadays, even the greater public is aware of this threat vector. In terms of numbers, ransomware has been accounted to attribute for up to 20\% of all cybercrime~\cite{malware-358}. Economically speaking, the average cost of a ransomware attack has been estimated in the range of 1 to 8 Million USD~\cite{ransomware}. While ransomware can be seen as a business-endangering attack, there are even specific breaches where the unavailability of ransomware has led to even more severe damages. For example, in the case of the Brno University Hospital in the Czech Republic, the deployment of ransomware led to the redirection of patients and the inability to conduct urgent surgical interventions~\cite{brno, ransomware-health}.

The constant evolution of ransomware attacks, in terms of attack behavior and sophistication, has stimulated research on how to defend against it. Nowadays, thousands of papers address different areas of the cyber kill chain of ransomware. For example, industry leaders such as IBM~\cite{ransomware-ibm} name threat detection as one of the main pillars of defense. Here, AI-driven solutions are vital to detect the constantly changing attack behavior~\cite{ransomware-ibm}. Indeed, numerous papers have demonstrated high accuracy in detecting ransomware, many of them achieving 99\% accuracy~\cite{icc, specforce, intelligent-beh-iot, entropy_file_detection}. 

Looking at the broad coverage of ransomware detection in current research, one could reason that the ransomware threat can be disregarded in light of intelligent intrusion detection systems. However, from an incident response perspective, the current state of the art presents multiple challenges. First, even though ransomware is an important threat, on a system level, ransomware is rare to be encountered since the large majority of runtime would be spent in a benign state. Due to this low base rate, even a high accuracy in detecting ransomware may lead to many false positives~\cite{dosanddonts}, which must be resolved through incident response. This is costly since a single false alert requires up to 30 minutes of active human investigation~\cite{alert-fatigue}. 

Secondly, ransomware defense is mainly centered around recovery~\cite{ransomware-ibm}. As such, most systems are not resilient enough to prevent ransomware but instead focus on recovering data. In general, research in intrusion detection does not always investigate or discuss the actual implications of the defense that it enables~\cite{dosanddonts}. Therefore, there is a clear opportunity to investigate the usefulness of AI-driven detection in conjunction with an integrated, reactive defense that can mitigate ransomware in a resilient way while demonstrating the portability of the models and their application in realistic settings, including novel AI-based evasion methods~\cite{globecom}.

To address the previously mentioned challenges, the work at hand presents the following contributions:
\begin{itemize}

    \item The design and implementation of an integrated detection and mitigation framework for ransomware in Linux-based systems. The proposed framework uses the file system abstraction present in the operating system as a target level to implement this functionality. Thus, a file system composed of an overlay and underlay file system is presented. The detection uses file system-related system calls to extract data on a process level. By processing a variety of features extracted during a pre-defined duration, individual processes can be classified as malicious or benign. For mitigation, several novel and baseline defense strategies, including deceptive and defensive techniques, are designed for various workloads that are differentiated by data sensitivity and latency criticality.

    \item The instantiation and deployment of the framework on a Raspberry Pi acting as an FTP server affected by three ransomware samples. Here, the detection system's delay and accuracy performance is assessed when presented with known and unknown malware while running a benign workload in parallel.

    \item To measure the effectiveness (\ie security guarantees) and efficiency (\ie the resource consumption), a pool of experiments have been performed. Here, a virtualized testbed using a fast storage underlay system is used to measure the usefulness of reactive defense from a mitigation perspective. Specifically, eight ransomware samples were retrieved and tested against seven different defense scenarios. The resulting 48 experiments assess the number of bytes lost and the resource consumption when deploying each ransomware. Several experiments are conducted to assess the usefulness of the different defense strategies when running benign workloads, which are oriented towards specific use cases (\eg server administration, sensor persistence). Based on this analysis, recommendations are made regarding selecting a defense strategy given the data criticality and latency requirements of the benign workload.
    
\end{itemize}

The remainder of this work is structured as follows. Section~\ref{sec:related} reviews related work dealing with dynamic ransomware detection systems. While Section~\ref{sec:framework} presents the design of the proposed framework, Section~\ref{sec:implementation} shows its implementation details. Then, Section~\ref{sec:setup} introduces the testbeds and ransomware samples used to validate the framework. Section~\ref{sec:experiments} evaluates the framework detection performance and consumption of resources in the previous scenario. Furthermore, the applicability of the defense methods against different sets of requirements is analyzed. Finally, Section~\ref{sec:conclusion} presents conclusions of this work and outlines future areas of research in ransomware mitigation.

\section{Related Work}
\label{sec:related}
This section reviews the literature combining ransomware detection and mitigation. 
Neglecting the adversarial context of detection systems is a common pitfall~\cite{dosanddonts}. Thus, ransomware detection systems are analyzed from a defense perspective and discussed based on their applicability.

\textit{Cryptolock}~\cite{cryptolock} safeguards against Windows-based Ransomware by warning users when there is a possibility of encryption occurring on the system. The system assesses each process through a reputation score, which indicates the level of suspicion associated with that process over time. 
Additionally, the system employs a similarity hash function called \textit{sdhash}~\cite{sdhash} to gauge the likeness between the original version of a file and its updated version after a write operation. 
Since the score is computed after the file is written, data loss prevention is only realistic after the detection duration.

Another approach that leverages the notion of entropy is presented by~\cite{entropy_file_detection}. In their approach, the ransomware detection system computes the entropy based on the file name suffix. For comparison, a backup system is used to compute and compare the entropy of user data. In an optional step, multiple Machine Learning (ML) models are available if the system load allows their application in terms of system resources. 
The authors stress that this computation has to be applied by categorizing data based on the types of files that are accessed. While the authors present highly promising results with respect to the detection of malware, they do not directly assess the effectiveness or efficiency in a real-world defense scenario. Furthermore, the authors only outline that file recovery is a conceptually compatible strategy. 

Since many ransomware instances not only encrypt data but also communicate over the network (\eg for key exchange, operational control), \textit{RansomSpector}~\cite{ransomspector} monitors both the file system and network traffic to provide improved accuracy. 
The proposed monitoring component captures the system calls via a hypervisor and checks whether the system call is related to the file system (\eg OPEN, LINK, WRITE) or network activity (\eg CONNECT, BIND). If it is, the system call is sent to the Detector, which then performs pattern matching. If both file system and network operations match the malicious pattern, the process is identified as malicious. 
RansomSpector has been evaluated on Windows 7 to establish the detection rate and performance consumption, not the defense effectiveness. Due to the alert-based mitigation strategy, files may not be preserved. Thus, substantial management efforts may need to be dedicated to resolving false and true positives (\ie file recovery).

\setlength{\tabcolsep}{4pt}
\begin{table*}[t]
\footnotesize
    \caption{\label{tab:categorization} Categorization of Surveyed Related Work}
    \centering
    \begin{tabular}{@{}llccllll@{}}
        \toprule
        \textit{Work}  & \textit{Mitigation} & \textit{Prevent} & \textit{Detect} & \textit{Platform} & \textit{Evaluation Metrics} & \textit{Data Collection} \\\midrule
       \cite{cryptolock} 2016  & Alerts, Process Termination & \xmark & \checkmark & Windows &  \textsc{D|M} & User Data \\
       \cite{shieldfs} 2016 & Shadow Drive Buffering, Discard Writes & \checkmark & \checkmark & Windows & \textsc{D|M|P|U}  & IRPLogger \\
       \cite{redemption} 2017  & Buffers, Termination & \checkmark & \checkmark & Windows & \textsc{D|P|U} & I/O System \\
       \cite{entropy_file_detection} 2019   & Backup Recovery & \xmark & \checkmark & Backup Systems &  \textsc{D} & \xmark \\
       \cite{ransomspector} 2020 & User Notification & \xmark & \checkmark & KVM/Windows & \textsc{D} & System Calls \\
       \cite{icc} 2023  & Trap Directories, File Renaming & \xmark & \checkmark & Linux  & \textsc{D|P} & Performance Metrics \\
       \cite{mtfs} 2023  & MTD: Trapping, Delays & \xmark & \xmark & Linux  & \textsc{M|P} & Performance Metrics \\
       \cite{fesad} 2024  & Isolation, Process Termination & \xmark & \checkmark & Windows Sandbox  & \textsc{D} & Windows API Calls \\
       \cite{xran} 2024  & Explainable Alerts & \xmark & \checkmark & Windows Sandbox  & \textsc{D|P} & Windows API Calls \\

        This   & Process Termination, Delaying & \checkmark & \checkmark & Linux  & \textsc{D|M|P|U} & File System Calls \\
        & Deceptive Modifications, Tracking & & &  &  &  &  \\
        \bottomrule
    \end{tabular}
    \label{tab:summaryRelatedWork}
       \textsc{D}=Detection Evaluation, \textsc{M}=Mitigation Evaluation, \textsc{P}=Performance Analysis, \textsc{U}=Usability Impact Analysis
\end{table*}
\setlength{\tabcolsep}{6pt}


\textit{Redemption}~\cite{redemption} stands out as a newly proposed system capable of detecting and preventing ransomware. It achieves this by intercepting write operations, redirecting them to mirrored files, and preserving the original files. The implemented kernel intercepts file system operations calculates entropy ratios, utilizes File Content Overwrite, and considers access frequency to identify suspicious activity. 
To evaluate their system, Windows was chosen as the only target. Using this testbed, they assessed the detection performance and the resource overhead incurred by the system. Furthermore, they conducted usability experiments. While the results are promising, it is not clear whether the defense metrics were measured or whether they were inferred from the mitigation performance. Furthermore, \cite{globecom} demonstrated that explicit feedback to ransomware can be used by an adaptive attacker since the solution is not deceptive. 

\textit{ShieldFS}~\cite{shieldfs} is a self-healing file system that adds ransomware protection capabilities to the Windows operating system. For the data collection, an I/O file system sniffer has been developed that incorporates additional information, namely entropy, process identifier (PID), and timestamp. 
When trained classifiers report malicious behavior, the file system discards written copies and terminates the process. 
To assess the effectiveness of this approach, virtual machines were used for experimental deployment. Aside from assessing the classification, it was assessed how well the system can recover the files. Aside from promising detection and prevention results, several limitations are present. First, the system was only assessed for the Windows platform. 
Furthermore, DoS attacks could be executed on the buffers, and the termination signal could be used for adaptation~\cite{globecom}.

\cite{icc} presented an ML-based detection system using performance metrics gathered from the Linux kernel. Among other malware, ransomware was identified with high accuracy. Different Moving Target Defense (MTD) techniques can be deployed after successful detection.  
As shown during experiments with real malware, these techniques allow the ransomware to be terminated at some point. Nevertheless, reactive deployments lead to data loss, while proactive ones were considered wasteful in terms of resources.  

Novel approaches in ransomware detection acknowledge the evolutionary nature of ransomware, hence the necessity for a detection system to keep up with the changing nature of ransomware. \textit{FeSAD}, a recently proposed framework, was demonstrated to apply binary classification for this context successfully. While the work confirms the effectiveness of using Windows API calls and reasons about the applicability of potential mechanisms such as system isolation or task termination, no preventative measures are proposed~\cite{fesad}.

\cite{mtfs} provides a virtualization of the MTD-based defense from~\cite{icc}, leading to multiple defense methods. First, the recursive (\ie infinite) directory tree defense is implemented in the file system, where a directory listing is extended by a specific directory. If this directory is listed, the directory itself is returned, leading to an infinite trap if one were to apply a depth-first directory traversal. Furthermore, operations can be cheaply delayed, and it is possible to obfuscate the file suffixes and the identifying magic bytes. 
The effectiveness is analyzed in a reactive setting, showing that multiple ransomware samples can be completely prevented. Importantly, this achievement relies on the traversal strategy of the samples employed. Moreover, no assessment of the impact on benign workloads has been performed. 

Arguing that existing models for ransomware detection lack explainability, \cite{xran} propose \textit{XRan}, a Convolutional Neural Network-based model. The authors leverage multiple dynamic and static sources (\eg Windows API calls, linked libraries) to achieve local and global explainability. While the authors add an effective and explainable detection method to the growing body of literature, the effectiveness of the approach in a defense scenario is not tested, and no discussion on portability to other platforms is present.

Table~\ref{tab:summaryRelatedWork} summarizes the previously analyzed contributions in terms of their mitigation, prevention, and detection strategies, the platform they target, how data is being collected, and the approach to evaluating the work. In conclusion, despite the advances in OS-level ransomware detection and mitigation, the following challenges are still open. First, only limited experience is drawn from platforms targeting Linux-based operating systems. Secondly, most ransomware mitigation approaches informed by a reactive deployment scenario terminate the encryption behavior. However, the understanding of how to stop and prevent ransomware during execution is sparsely covered. Similarly, the usefulness of a detection system is not brought into the context of an actual defense scenario. As a fourth challenge, all ransomware mitigation approaches provide feedback to the attacker (\eg file recovery, process termination). As showcased in~\cite{globecom}, this threat model may not be realistic anymore. In summary, there is no work applying a reactive ransomware defense system that deceptively prevents ransomware on Linux-based systems, especially when considering a broad set of real-world samples in realistic execution settings.

\section{Framework Design}
\label{sec:framework}

The section presents a framework for integrated detection and mitigation of ransomware in autonomous Linux-based devices by applying both aspects from a file system abstraction level. More in detail, it shows the details of the framework architecture and the threat model that is subsumed under the threat vector commonly termed "ransomware."

\subsection{Ransomware Threat Model}
Although the term ransomware is often used unanimously during threat modeling, there exist various flavors.  

The most aggressive ransomware variant is \textit{crypto-ransomware}, which traverses the file system of the target and potentially mapped file systems (\eg NFS, Samba shares). For each of the files of interest (usually a subset of all files), an encrypted copy is produced in lieu of the original file. Thus, the availability of the original data is threatened, for at least the time it takes to restore the data from a backup. In the worst case, all data can be lost if victims are not able to retrieve the encryption key~\cite{cryptoransomware}.

When the key cannot be retrieved by means of law enforcement, victims consider the payment of the ransom that is demanded by the attackers, although it is not clear if the key is actually released. Without actually encrypting data, \textit{scareware} tries to trick users into believing that their data has been compromised, although it may not have actually been breached by the attackers. In that sense, one could reason that this threat vector is less impactful~\cite{scareware}.

More similar to crypto-ransomware is \textit{locker malware}, which tries to make data or functional assets unavailable by other means than file encryption. For example, a specific lock screen may be installed by the malware to lock the user out of the device~\cite{locker}.

\textit{Ransomware-as-a-service} cannot necessarily be differentiated by its damage or breach function but rather describes the business model and degree of sophistication of the threat actors. Here, a ransomware strain can be bought from the attackers, which may also offer to take care of other aspects such as payment processing~\cite{raas}.

Finally, \textit{extortion}-based ransomware aims to increase the chances of paying the ransom. Here, one or more of the previously described attack vectors (\eg file encryption or deletion) are combined with the leakage of sensitive data so that victims can be pressured into paying the ransom to avoid publication of the data~\cite{extortion}.

With all the previously defined damage methods, a wide array of infection methods are possible. For example, spear phishing is a common approach the gain initial access to the victim's infrastructure~\cite{extortion}. For the threat model considered in this work, the infection method is not ignored. Thus, it is assumed that the ransomware has user-space access to a Linux system and that it has sufficiently elevated privileges to be able to read, write, and delete files. Furthermore, it is assumed that the attacker can execute binaries and scripts on the host. With respect to the damage function, crypto-ransomware is considered due to its aggressive nature and its prevalence in conjunction with other behaviors.

\subsection{Architecture}

\begin{figure*}
    \centering
    \includegraphics[width=0.9\linewidth]{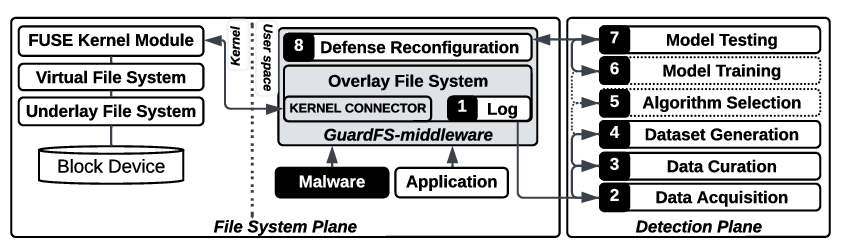}
    \caption{File System-based Framework Architecture}
    \label{fig:architecture}
\end{figure*}

To defend against the previously described ransomware attack model, the framework shown in \figurename{}~\ref{fig:architecture} is proposed. The framework is designed in a distributed manner, with two planes separating concerns and responsibilities. This allows that computationally weaker devices (\eg resource-constrained devices, mobile devices) can only implement the defense and data creation components, with an external node running all components related to the detection. However, a computationally capable device could run all components in the same execution environment. As presented in the architecture, the following two planes take care of two functionalities.

\begin{enumerate}
    \item \textit{File System Plane}. It is a fully functional overlay file system in charge of servicing any file system-related system call. The behavior of the call handler depends on the defense configuration. Secondly, the file system acts as a data logger for the detection system.
    \item \textit{Detection Plane}. It provides intelligence in the device for reactive mitigation. Input data can be collected from the file system plane. Due to its positioning in user space, additional data sources can be integrated.
\end{enumerate}

\subsubsection{File System Plane}
The most important task of the file system plane is to act, as the name implies, as a file system. Since it is conceptualized as an \textit{overlay file system}, this means that it will receive system calls from an abstract representation of a file system (\ie the \textit{virtual file system} running in the kernel). The overlay file system must handle the system call and return the appropriate response to the calling process. For example, when a hypothetical process writes to a file, the following system call is invoked in the overlay file system.

\definecolor{mGreen}{rgb}{0,0.6,0}
\definecolor{mGray}{rgb}{0.5,0.5,0.5}
\definecolor{mPurple}{rgb}{0.58,0,0.82}
\definecolor{backgroundColour}{rgb}{0.95,0.95,0.92}
\lstdefinestyle{CStyle}{
    backgroundcolor=\color{backgroundColour},   
    commentstyle=\color{mGreen},
    keywordstyle=\color{magenta},
    numberstyle=\tiny\color{mGray},
    stringstyle=\color{mPurple},
    basicstyle=\footnotesize,
    breakatwhitespace=false,         
    breaklines=true,                 
    captionpos=b,                    
    keepspaces=true,                 
    numbers=left,                    
    numbersep=5pt,                  
    showspaces=false,                
    showstringspaces=false,
    showtabs=false,                  
    tabsize=2,
    language=C
}
\begin{lstlisting}[style=CStyle]
ssize_t write(int fd, const void buf[.count], size_t count);
\end{lstlisting}
Now, the overlay file system can invoke the same system call on the \textit{underlay file system}, which can be any file system mounted on the Linux machine that operates on the underlying storage. 
To do so, it sends the same system call, passing the file handle \texttt{fd}, buffer \texttt{buf}, and a number of bytes written \texttt{count}. Finally, it returns the \texttt{count} back.

With this behavior, the overlay file system forwards calls to the file system, which is simply the existing file system implementation in a Linux operating system. However, the actual execution of the call would depend on the \textit{defense reconfiguration}. This configuration describes how to intercept system calls. For example, one configuration could involve the lookup of the calling process by process identifier (PID) and ignore the system call.
To design a defense configuration, it is important to consider the ransomware threat model. For example, for ransomware that overwrites the target file, the \texttt{write()} system call behavior can be modified. However, other crypto-ransomware will simply create a new file and delete the target file -- thus, the \texttt{unlink()} system call is considered. The actual behavior of the defense can be made more granular, depending on the output of the detection system. 

The second task of the file system is to provide data to the \textit{Detection Plane}. This is especially critical when mitigation and detection should be aligned in a way where they enable not only the detection of damage done but also the prevention thereof. For example, a detection system could easily detect encrypted files based on resource consumption (\ie ongoing encryption)~\cite{icc} or based on activity on the files in the normal file system. However, this would detect damages already done to the system. Thus, the (overlay) file system presents a unique opportunity, as the system calls received and the data passed in them is "the last mile" before the damage is written to disk. Nevertheless, data collection is optional, and proactive defenses could be implemented, too.

There are dozens of types of system calls from which data can be captured. Thus, the first dimension is the type of system call. Similarly, many system calls are parametrized by flags, which can be captured. Next are the file paths and file descriptors. Here, the file suffix can be an important dimension since ransomware differs by the subset of file types they target~\cite{mtfs}. It is important to consider that only calls such as \texttt{open()} actually give access to the file path, whereas others use the internal representation of a descriptor. Another important dimension is buffers, such as the ones passed by \texttt{write()} and \texttt{read()}. The collection of the buffers can lead to high resource consumption. One approach is to transform it into the entropy, which measures the randomness. For example, Equation \ref{entropy} computes the Shannon Entropy.
\begin{equation} \label{entropy}
H = -K\sum_{i=1}^{m}p_i  log(p_i)
\end{equation}

\subsubsection{Detection Plane}
The \textit{Detection Plane} provides all necessary capabilities to implement ML/DL-driven reactivity to reconfigure defenses. Thus, this plane houses all data analysis-related aspects, starting from \textit{data acquisition} all the way to the actual \textit{model evaluation}. As presented in Figure~\ref{fig:architecture}, the model training and algorithm selection steps are only considered during the implementation (or its continuous refinement) of the proposed platform.

The main consideration in the \textit{Data Acquisition} step is how data can be pulled from the file system. Thus, the actual business logic is informed by the behavior of the data producer. For example, if logs are provided by the file system plane in a continuous stream, they should be consumed in a stream-based manner. If logs are only written in a periodical manner, regular checks are performed to read the new logs. Logs from a file system, especially based on system calls, can contain many elements per slice.
Furthermore, the size of the logs can vary greatly, with some logs potentially being empty when no process accesses the file system during the monitoring cycle.

This synchronization and preliminary buffering must be considered in the \textit{Data Curation} step, along with the selection of features and the processing of computed features. At this stage, data from the file system can be correlated with other data. For example, for each process that was involved with file system activity, data about its resource consumption can be extracted from the process monitor. For example, \cite{some-huertas-paper} presents an overview of kernel events and performance metrics that can be extracted. Here, many pitfalls must be considered, as outlined by~\cite{dosanddonts}. For example, spurious correlation between the features and the class it represents must be avoided. To finalize the creation of a dataset, the \textit{Dataset Generation} step persists the data and splits it into a training and evaluation dataset. 

During the instantiation of the framework, the next step is the \textit{Algorithm Selection}, where different algorithms for the classification of the data are evaluated. The most appropriate one is selected for the \textit{Model Training} step, where the data is fed to the algorithm, producing one or more models based on the hyperparameters defined. Finally, the \textit{Model Testing} stage applies to the creation of the framework and to its continuous operation. For the latter, new vectors are continuously evaluated to produce a classification of the file system activity. Importantly, the output must be published to inform a closed loop between the two planes. After all, successful detection of malicious behavior is only effective if the right action is taken in a time-effective manner. Thus, the classification results are published, at least for results that evaluate to malicious behavior. For example, the identifiers of malicious processes can be written to a file log, queue, or socket so that the file system plane can reconfigure the defense behavior.

\section{Framework Implementation}
\label{sec:implementation}
With the conceptual elements in the framework introduced, this section presents a specific instantiation of the framework, consisting of an overlay file system with multiple defense configurations and a ML-based binary classifier for reactive deployment of the configurations.

\subsection{File System Plane}\label{fs}
The file system plane is implemented as an overlay file system by leveraging bindings to the \textit{FUSE}~\cite{fuse} library from the \texttt{Go} programming language~\cite{fuse-go}. Essentially, each system call destined to the subtree of the file system under the mount point can be hooked into, effectively overwriting the normal behavior. Thus, the overlay system is achieved by receiving system calls, adapting the parameters passed, and issuing a new system call. For example, in an \texttt{open()} system call, the file path of the file to open or create is passed. Here, the path is changed to reflect the path in the underlay, a new \textit{open()} system call is executed, and the results are passed to the caller.

To gather data about malicious and benign processes, a separate threat is collecting data for a number of seconds before writing it to an output stream that is also persisted in the underlay file system. Three different intervals (\ie 1, 5, and 10 seconds) were evaluated for buffering. In theory, longer buffers should present richer behavioral data~\cite{earlydetection-survey} but at the cost of delayed detection and higher memory constraints. Based on a preliminary analysis of three ransomware samples, only the \texttt{read()} and \texttt{write()} operations and their parameters were considered for persistence. Due to the high number of system calls produced by the aggressive ransomware behavior, the following dimensions are recorded: \1 the process identifier (PID) is recorded to distinguish between processes. For \texttt{write()} calls, \2 the Shannon Entropy of the buffers passed is calculated. Furthermore, \3 a timestamp is recorded so that additional metrics can be computed. Furthermore, \4 the file path, including the file name and suffix, are recorded.

Besides collecting data and routing system calls, the file system plane implements four different defense configurations that can be reactively and selectively applied. This means that two processes could access (\eg write to or read from) the same file. However, only the benign one will be forwarded to the underlay behavior, while the malicious one will be deceived or mitigated.

\subsection{Defense 1: Killing Processes (PKILL)}
\begin{figure}[h]
    \centering
    \includegraphics[width=\linewidth]{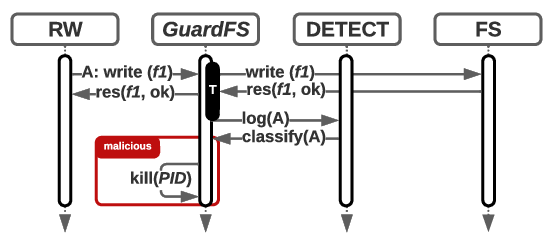}
    \caption{\texttt{PKILL} Defense}
    \label{fig:def-pkill}
\end{figure}

The first technique is the conceptually simplest one, which has already been explored in the literature~\cite{cryptolock, fesad}. As such, it should also serve as a comparison baseline for other defense configurations. Furthermore, it is implemented in the overlay file system to add the novel element of delaying any modifying system calls until a time $T$ has expired. This duration shall allow enough time to gather data about the process, classify its behavior, and decide how to react. As presented in Figure~\ref{fig:def-pkill}, \texttt{PKILL} invokes a process termination through the operating system based on the PID if the file system receives the signal of it being malicious. Any explicitly benign behavior is forwarded to the underlying file system while returning this response to the calling process of the system call.

\subsection{Defense 2: Obfuscating Responses (OBF)}
While the previous defense may be successful in deterring certain types of ransomware, it inadvertently shows weaknesses. First, it does not protect against upcoming damage, and a controller could redeploy the malware. Secondly, as shown in a recent research work~\cite{globecom}, a limitation of this type of defense is that it presents an explicit trigger to the attacker. Thus, an attacker will be able to learn precisely under which circumstances the ransomware was detected, potentially leading to an adaptation of the behavior. Thus, \texttt{OBF} presents a defense that does not give an attacker explicit feedback while protecting against loss of data. 

\begin{figure}[h]
    \centering
    \includegraphics[width=\linewidth]{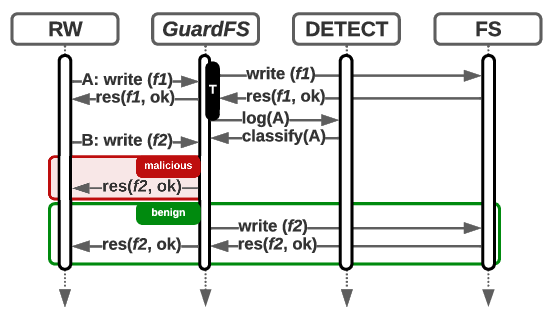}
    \caption{\texttt{OBF} Defense}
    \label{fig:def-obf}
\end{figure}
As shown in Figure~\ref{fig:def-obf} any operations of unknown processes are immediately forwarded to the underlay systems. During each time period $T$, data is collected about all processes. After evaluating the current buffer of logs, malicious processes are continuously given access to the underlay file system. However, for malicious processes, the PID is cached in the file system, and any damage-inflicting system calls are decepted. Thus, a \texttt{read()} system call is still granted; however, \texttt{write()}, \texttt{rename()}, and \texttt{unlink()} (\ie deleting a file or directory) are ignored in a special way. No actual system call to the underlay is executed. However, a falsified response is crafted to let the caller believe it was executed. For example, a \textit{write()} system call passes the file descriptor and a buffer. The caller expects a single number that indicates how many of the bytes in the buffer were written. Thus, this defense configuration heuristically waits a small amount of time and then responds with the length of the input buffer, leading the caller to believe that all data was successfully written.

\subsection{Defense 3: Delaying and Obfuscating Responses (DEL+OBF)}
The previously described defense (\ie \texttt{OBF}) applies response obfuscation to all file system operations that would lead to changes in the underlay system if they were issued from a process that was previously classified as malicious. \texttt{DEL+OBF} is designed for use cases where data is even more sensitive, but latency is not critical. For example, a data collector for a long-term heart monitor may constitute highly sensitive data. However, the latency might not be critical, especially if the sensor already transmits collected data in batches (\eg once per hour). In such a case, it only matters that the data is eventually persisted, but since there are no immediate read operations, a short delay can be tolerated.
\begin{figure}[h]

    \centering
    \includegraphics[clip,width=\linewidth]{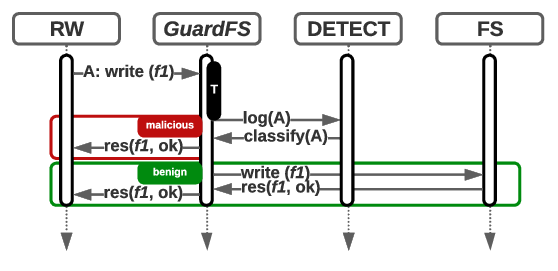}
    \caption{\texttt{DEL+OBF} Defense}
    \label{fig:def-del-obf}
\end{figure}

Thus, this technique implements the same defense for maliciously classified processes (\ie obfuscate system call responses). However, as shown in Figure~\ref{fig:def-del-obf}, all system calls of all processes are blocked for the duration of the timer $T$. By adhering to this timer, the situation where a system call is responded to without knowing if its payload is malicious or not does not arise. In theory, this is done at the expense of the buffering requirements represented by the following equation, which have to be accommodated in the memory of the device.

\begin{equation}
  min(\delta, \epsilon) + min(\delta, \beta) \times T
\end{equation}

Intuitively, the memory requirements are calculated for the duration of the buffering. Then, the duration is multiplied by the effective encryption rate, which is defined by the encryption rate $\epsilon$, controlled by the ransomware, and by the disk or file system throughput $\delta$, which limits the speed of the ransomware. Although the ransomware encryption rate may not seem useful, it is an important construct, as will be presented in the experiments, since highly sequential ransomware will be limited in terms of encryption rate due to the buffering itself since they do not implement an asynchronous (\ie non-blocking) traversal and encryption. Analogously, the throughput for all malicious processes is considered by the factor of $\beta$, since the buffering has to be performed for these processes, too. Thus, it is important to consider that buffering and blocking are applied to all processes, since the trustworthiness is not known beforehand and it is not considered for the full process lifecycle.

\subsection{Defense 4: Tracking Processes (TRACK)}
In terms of design, \texttt{DEL+OBF} should present the highest security guarantee, while \texttt{OBF} presents the lowest latency of benign processes. The final configuration \texttt{TRACK} tries to maintain state information about long-running processes and discriminate between malicious and benign processes by looking at the calling PID.

\begin{figure}[h]
    \centering
    \includegraphics[trim={0.15cm 0cm 0.2cm 0},clip, width=.9\linewidth]{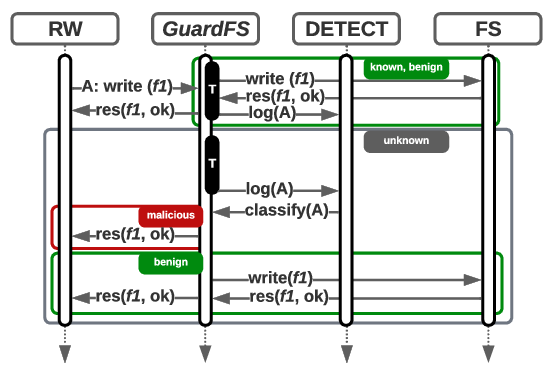}
    \caption{\texttt{TRACK} Defense}
    \label{fig:def-track}
\end{figure}

As presented in Figure~\ref{fig:def-track}, unknown processes receive the same behavioral response as all processes do in \texttt{DEL+OBF} -- they are responded to in a blocking manner. Thus, for time duration $T$, their activity is collected by the logger and classified by the detection system. Only then is a response created. The response behavior again follows the same obfuscating behavior for malicious processes while it forwards system calls for benign processes. This state information is then tracked for subsequent calls to the file system by adding the PID to a hash table.

For known benign processes (\ie for hits in the hash table), the system calls are immediately forwarded to the underlay file system. However, that does not mean that their behavior is not monitored. In an opposing way, their behavior is tracked by a different \texttt{monitoring} thread for a time duration $T$, after which the classification is evaluated, and the hash tables are updated. In the case of a benign process that at some point exhibits malicious behavior, the average data loss would be represented by the following Equation (assuming that malicious behavior would be detected).
\begin{equation}
  min(\delta, \epsilon)  \times T \times 0.5
\end{equation}

For processes that are known to be malicious, the file system will immediately invoke the obfuscating responses. This is an important factor to consider since, in this configuration, ransomware may only be blocked for one monitoring iteration. Afterward, responses are obfuscated and sent back at high throughput without blocking the calling process. In parallel, data is still continuously collected, even though the process was classified as malicious.

\subsection{Detection Plane}\label{sec:detection-impl}
Due to the pure user-space implementation of the file system plane described in~\ref{fs}, several approaches for detecting ransomware are feasible. For example, kernel metrics, hardware counters, system calls, or performance measurements could be used. However, to implement a pure file system approach, only related system calls are used. Thus, in the monitoring thread that implements the \textit{Detection Plane}, a dataset is curated from continuously reading the system calls and aggregating them to buckets. The bucketing is applied so that for each time slice, the metrics in Table~\ref{tab:dataset} are aggregated.

To create the malicious and benign behavior, three ransomware samples were executed. More specifically, \textit{RansomwarePoC}, \textit{DarkRadiation}, and \textit{roar} (see Section~\ref{sec:malware-samples}) were deployed by directly giving them shell access. For benign behavior, an FTP server was deployed on the same Raspberry Pi device (see Section~\ref{sec:setup-rasp}), and the load was generated using the Apache JMeter~\cite{jmeter} stress testing suite. By configuring multiple client threads, a high load was placed on the server, consisting of reading and traversing the directory structure. The files deployed on the device used a broad set of file types provided by~\cite{govdocs}. Another subset from the corpus was used to upload it to the device so that write operations with high entropy (\eg ZIP archives, JPG images) are also present in benign behavior. Under these conditions, normal data was collected for 258 minutes ($\approx$ four hours). Data collection involving encryption by each ransomware sample and benign behavior spanned several hours. \textit{roar} did not achieve full encryption in that time and was, therefore, terminated after roughly two hours. Due to the low encryption activity of \textit{roar}, only 1.7\% of the collected system calls are labeled as malicious in the resulting dataset. For \textit{DarkRadiation}, 13.9\% of all system calls were labeled as malicious, and for \textit{RansomwarePoC}, 9.2\% were malicious calls.

\begin{table}[pos=b]
    \centering
    \caption{Shape of Aggregated Dataset}
    \begin{tabular}{@{}ccccccc@{}}
        \toprule
        \textit{time} & \textit{writes} & \textit{reads} & \textit{pid} & \textit{e\_min} & \textit{e\_mean} & \textit{e\_max} \\\midrule
        20 & 131 & 102 & 232 & 7.86 & 7.86 & 7.86 \\
        20 & 73 & 2 & 533 & 7.94 & 7.95 & 7.96 \\
        \toprule
    \end{tabular}
    \label{tab:dataset}
\end{table}

Once data was collected, the datasets that included the presence of different behaviors (\ie benign or one of the three ransomware samples) were aggregated into buckets of 2, 5, and 10 seconds. Furthermore, time-sensitive or leaking features (\eg file extensions, paths, time, PID) were removed. Ultimately, each row in the dataset holds for each process the number of operations per system call type (\eg read, write, rename) and the minimum, maximum, and mean average of the buffers. Finally, each dataset was split, where 80\% was to be used for training and 20\% for testing.

To implement the model training component using ML, two approaches were followed. First, an aggregated dataset was used to create one global model, where all ransomware data was labeled as malicious and the remaining data as benign. In the second approach, three models were trained for each bucket configuration. Here, only two out of three ransomware samples were included in the training data to analyze whether unknown ransomware behavior can be detected based on the behavior of other samples. Of course, for anomaly detection, only benign data was used.

Thus, for both approaches and all three bucketing configurations, a model was created using three algorithms: Random Forest Classifier, Logistic Regression, and Isolation Forest~\cite{scikitlearn}. As shown in the subsequent experiments, all classifiers achieved high accuracy in classifying the malicious behavior using the default parameters provided by \textit{scikit-learn}, so no parameter tuning involving a validation split was used.

\section{Validation Scenarios}
\label{sec:setup}

To assess the effectiveness and efficiency of the described prototype, it was deployed in multiple scenarios, ranging from single-board computing to a container-based testbed. This section describes the configurations and related artifacts that were used in the experiments.

\subsection{Single-board Computing}\label{sec:setup-rasp}
Raspberry Pi devices are implemented as a system on a chip, making them a low-cost computing platform that can be used in a variety of use cases. Due to them being exposed to ransomware in the past, they fit the threat model described in this work. Furthermore, since they are considered resource-constrained devices (either in terms of computational resources or management capabilities), they present an excellent test bed that allows one to experiment with the effectiveness of the platform. Specifically, a Raspberry Pi 4B with 2 GB of memory was used to run an FTP server at high load. On this device, experiments are performed to assess the capabilities of the detection system both in an online and offline experiment (\ie with data gathered on the device, but evaluated locally).

\begin{figure}[ht]
    \centering
    \includegraphics[trim={0.18cm 0.26cm 0.25cm 0.15cm},clip,width=.84\linewidth]{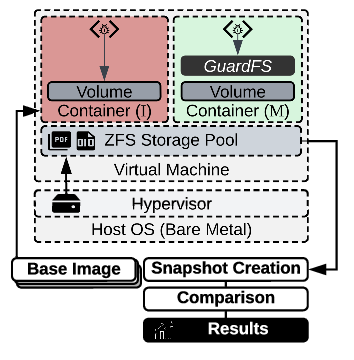}
    \caption{Virtualized Ransomware Testbed Architecture}
    \label{fig:testbed}
\end{figure}
\subsection{Virtualized Testbed}
\label{testbed}
While the deployment of real-world malware in a real device allows experimentation close to reality, it is tedious to gather data in a reproducible and scalable way so that numerous samples can be tested. Thus, the testbed shown in Figure~\ref{fig:testbed} is developed, where the hardware presents high-performance access to storage, by means of two NVMe storage devices and an AMD Ryzen 5700G processor running at 4.7 GHz with 64 GB of DDR4 memory. Experiments are executed in containers where a ZFS dataset is mounted. Optionally, a configuration of \solution{} can be mounted, too. Since the underlay resides on the ZFS dataset, snapshots can be easily created and compared to understand the effects.

\subsection{Malware Samples}\label{sec:malware-samples}
Aside from \textit{RansomwarePoC}, \textit{DarkRadiation}, and \textit{roar}, which have already been used during the implementation of the framework (\ie for the model creation step), a broad set of malware was obtained from malware databases and integrated into the testbed:
\begin{itemize}
	\item \textit{Babuk} is sophisticated malware,  whose source code was leaked. The ransomware is written in \texttt{golang}, targeting different platforms. Thus, the encryption module was extracted and compiled for the x86 platform~\cite{babuk}.
	\item \textit{Blackbasta} is a renowned ransomware-as-a-service enterprise. A leaked binary was obtained as an \texttt{ELF} file. After reverse-engineering it was discovered that it targets specific folder paths used by \textit{VMWare}. Thus, \solution{} is mounted on \texttt{/vmfs/volumes}~\cite{malwarebazaar}.
	\item \textit{Cl0p} is a strain of a famous group of attackers. A binary was fetched from MalwareBazaar~\cite{malwarebazaar}.
	\item \textit{Conti} operates as a service since 2020. A binary is available in MalwareBazaar~\cite{malwarebazaar}.
	\item \textit{DarkRadiation} is a ransomware that targets Linux-based systems. This sophisticated ransomware is implemented entirely in bash using Telegram for communication instead of a dedicated C\&C server~\cite{darkradiation}.
	\item \textit{GoCry} is an educational, open-source ransomware written in \texttt{golang}~\cite{gocry}.
	\item \textit{javaRansomware} is an educational, portable, open-source ransomware developed in \texttt{Java}~\cite{javaransomware}.
	\item \textit{lockbit} is another strain of the real-world ransomware that was operational in 2023~\cite{malwarebazaar}.
	\item \textit{lollocker} is an open-source ransomware straing using \texttt{bash} to orchestrate the encryption using \texttt{OpenSSL}~\cite{lollocker}.
	\item \textit{Monti} is a modified strain of existing real-world ransomware~\cite{monti}, which can be retrieved as an \texttt{ELF}-file on MalwareBazaar~\cite{malwarebazaar}.
	\item \textit{Ransomware-PoC} is a proof of concept open-source Python ransomware payload~\cite{ranssomwarepoc}. 
    \item \textit{roar} appears the most technically elaborate ransomware. Technically, it is an open-source adaption of \textit{RansomwarePoC}, aiming at diversifying the encryption behavior (\eg encryption algorithm, encryption speed) and optimizing the most stealthy operation by using Reinforcement Learning~\cite{globecom}.
  \end{itemize}

\section{Experiments}
\label{sec:experiments}

This section presents a pool of experiments that evaluates the performance of the proposed framework while detecting and classifying the ransomware families introduced in Section~\ref{sec:setup}. First, the detection plane is evaluated against the testing datasets to understand how well it detects ransomware samples in two cases. First, when the given ransomware sample is present in the training data, and second, when it is considered an unseen behavior. Next, the experiments move to a production scenario, considering the Single-board computing scenario, where the detection delay is contextualized in terms of files encrypted until detection. In the end-to-end experiments, the detection plane is integrated with the defense techniques. Here, the previously described virtualized testbed is employed so that the seven different configurations of the defense strategies are confronted with eight ransomware samples to compute the amount of data that is lost. Finally, the overhead of the detection and defense components on benign workloads is established, leading to a comparison with related approaches.

\subsection{Evaluating Test Datasets}
To assess the performance of the detection plane in isolation, data obtained in the Single-Board computing scenario is leveraged, considering multiple samples. The actual evaluation is carried out on the remote device that was used for training the models.

\subsubsection{Classifying Unseen Ransomware}
To understand how well the file system behavior classification performs for unseen malware samples, the datasets were combined into nine different combinations to train one model based on the Ransom Forest Classifier. Thus, one model was trained for each combination of the three different time windows (\ie 2s, 5s, and 10s) and combining two out of three ransomware samples (\ie leaving out either RansomwarePoC, DarkRadiation, and roar). These three samples (see Section~\ref{sec:detection-impl}) were chosen as their approach, and technical implementation provides a diverse sample set. 

Table~\ref{tab:Accuracy_rf} shows the accuracy for the different combinations of ransomware data and time window sizes. The random forest classifier performed best when aggregating the behavioral data into 5- or 10-second slices. In the 2-second time window, the accuracy is the lowest, although just a slight difference compared to other others (\eg 2.21\% difference in the worst case). This can be explained by the fact that, when aggregating into the 2-second window, not enough file system operations may be gathered at all times since ransomware samples cannot constantly encrypt at full speed, since they, like every other userspace process, may be interrupted by another process or blocked by I/O operations.

\setlength{\tabcolsep}{3pt}

\begin{table}[t]
\centering
    \caption{\label{tab:Accuracy_rf} Accuracy for unseen Ransomware - Random Forest Classifier}
    \begin{tabular}{@{}c| c c c c@{}}
        \toprule
        \textit{Time Window} & \textit{RansomwarePoC} & \textit{DarkRadiation} & \textit{roar}
        \\\midrule
        2s   & 99.92\% & 94.71\% & 99.43\% \\
        5s   & 99.91\% & 95.55\% & 99.65\% \\
        10s   & 99.90\% & 96.92\% & 100\% \\
        \bottomrule
    \end{tabular}
\end{table}

Even in the 5-second time period, the accuracy of detecting roar is lower compared to the 10-second time window; evaluating the 5-second time window multiple times over 10 seconds would likely provide comparable accuracy to the model with a longer window size.

For both RansomwarePoC and DarkRadiation, the accuracy in the 10-second window size is close to 100\%. Thus, using these models, it is possible to detect the two strains, even when data was collected from other samples.
For DarkRadiation, the accuracy increases with increasing window size, with the highest accuracy observed at 96.92\%. The fact that DarkRadiation is detected with a lower accuracy is surprising since it encrypts without trying to be stealthy and in fact, encrypts at the highest speed. The result can be explained by the fact that DarkRadiation is the only sample that uses pools of subprocesses to parallelize the encryption process. Thus, grouping data based on PID leads to multiple vectors for each time window for DarkRadiation, so certain features may not reflect the malicious behavior as robust as others, as it leads to different patterns in the dataset, leading to difficulties during the classification stage. Another contributing factor is that the average entropy of RansomwarePoC and Roar is 6, whereas, in the case of DarkRadiation, it is around 8 due to differences in encryption behavior and implementation. Nevertheless, the random forest classifier is relatively robust to these differences, as the accuracy is above 94\% for all time windows considered.

\subsubsection{Classifying Known Ransomware}
Even in the previously described setup with partial training data, most vectors can be accurately classified for all window sizes. However, more data is available in practice, given the breadth of available ransomware samples. Thus, assuming that the behavior can be generalized over multiple samples, a dataset is created using training data from all samples. Again, models are trained by aggregating the data into three window sizes. Furthermore, we compare two algorithms (\ie Random Forest and Logistic Regression) for classification. After concatenating all Ransomware and benign datasets together and splitting the data into 80\% train and 20\% test datasets, the respective models were trained. Both logistic regression and random forest classifier have shown high accuracy - close to 100\%, as shown in Table~\ref{tab:Accuracy_rf_lr_all_samples}.

\begin{table}[b]
\centering
    \caption{\label{tab:Accuracy_rf_lr_all_samples} Accuracy of Models Trained on Three Samples}
    \centering
    \begin{tabular}{@{}lllllllll@{}}
        \toprule
        \textit{Algorithm} & \textit{2s}  & \textit{5s} & \textit{10s} 
        \\\midrule
        Logistic Regression   & 99.53\% & 99.76\% & 99.87\% \\
        Random Forest Classifier   & 99.93\% & 99.97\% & 99.98\% \\
        \bottomrule
    \end{tabular}
\end{table}

\setlength{\tabcolsep}{8pt}


In summary, the Random Forest Classifier for a window size of 10 seconds presents the best results. However, for practical reasons, the 5-second variant may provide comparable accuracy while providing merely timely detection.

\subsection{Evaluating after Deployment}
To understand how effective and efficient the detection plane can be when running in a real device, the Single-Board Computing (see Section~\ref{sec:setup}) scenario was instantiated. The detection system and the FTP workload run in parallel for each sample. Then, the delay between malware deployment and detection is measured. Furthermore, it is computed how many files were successfully encrypted by the sample in that time. This contrast is especially important in light of stealthy malware samples, such as roar, that decrease the encryption speed in favor of appearing less aggressive.
\setlength{\tabcolsep}{4pt}
\begin{table}[pos=h]
        \centering
        \caption{\label{time_elapsed_detection} Detection Delay (5s model)}
        \begin{tabular}{@{}lcccl@{}}\toprule
            {\textit{Sample}} & {\textit{Min.}} & {\textit{Avg.}} & {\textit{Max.}} & \textit{Data Loss} \\\midrule
            RansomwarePoC &4 s   &6 s  &10 s & 18 Files -- 12.3 KB \\
            DarkRadiation &4 s   &8 s  &10 s & 35 Files -- 23.9 KB\\
            Roar &8 s   &19 s  &44 s & 3 Files -- 2.1 KB \\
            \bottomrule
        \end {tabular}\par
\end{table}
The results, shown in Table~\ref{time_elapsed_detection}, reflect that intelligent ransomware such as roar is, in fact, able to evade detection better than other strains. Overall, the maximum detection delay observed across ten iterations of the experiment was 44 seconds, while the minimum was 4 seconds. However, putting this into the perspective of the encryption speed, stealthy ransomware such as roar cannot encrypt as many files as the other samples since it uses periodic phases of hibernation, which explains the variations of detection delay. 

Most samples can be detected based on a few seconds of \textit{active} encryption. While the number of files lost appears daunting, it has to be emphasized that in this scenario, no defense mechanism is present. As will be shown by the subsequent experiments, the file system defense can save some of the data that the ransomware is encrypting until the classification is positive.


\subsection{End-to-end Experiment}
So far, the detection plane was developed using a small number of ransomware samples, leveraging strains that differ in implementation and purpose. As such, the previous experiments demonstrated detection performance in offline and online settings. However, measuring the usefulness of an AI-based detection system for mitigating cyberattacks must include the complexities of the defense behavior. For example, active mitigation advertently changes the device's behavior and, ideally, even the malicious behavior, as the goal is to interrupt, diminish, or prevent the behavior. Thus, if such a system considers only detection without mitigation, the actual performance can only be approximated.

Thus, a series of experiments are executed using the virtualized testbed presented in Section~\ref{testbed}, spanning all malware samples from Section~\ref{sec:malware-samples}. Concerning the detection plane, the same Random Forest-based model from the online test is deployed in the testbed. Then, in each round, one workload is considered to assess \1 the defense effectiveness as established by the number of bytes lost when ransomware is deployed, \2 the resources consumed by the ransomware and the defense platform, and \3 the impact of the defense platform on benign workloads.

\subsubsection{Defense Effectiveness}
One experiment per defense configuration and ransomware sample have been performed. To compare the performance of the malware, each sample is also deployed against a baseline strategy, where no defense is active. At the beginning of each experiment, the detection plane is executed, which monitors access to the file system in the background. If a process is classified as ransomware, the defense strategy, which is the subject of the experiment, is deployed. The ransomware is given enough privileges to directly execute any operations on the files in the home directory. All files are available through the overlay file system to account for any encrypted files in the experiments. 

\begin{algorithm}
\caption{Pessimistic Computation of Data Loss}\label{alg:loss}
\begin{algorithmic}
\Require \textsc{baseline\_files} $\neq nil$
\Require \textsc{snapshot\_files} $\neq nil$
\State  \textsc{snapshot\_checksums} $ \gets []$

\State \textsc{files\_modified} $\gets []$
\State \textsc{bytes\_lost} $\gets 0$

\State \textsc{ptr} $ \gets 0$

\While{\textsc{ptr} $<$ \textsc{snapshot\_files}.length()}
\State \textsc{f} $\gets$ \textsc{snapshot\_files[ptr]}
\State \textsc{snapshot\_checksum} $\stackrel{+}\leftarrow$ sha256sum(\textsc{f})
\State \textsc{ptr} $\gets$ \textsc{ptr}$+1$
\EndWhile
\While{\textsc{ptr} $\leq$ \textsc{baseline\_files}.length()}
\State \textsc{f} $\gets$ \textsc{baseline\_files[ptr]}
\State \textsc{checksum} $\gets$ sha256sum(\textsc{f})

\If{\textsc{checksum} is not in \textsc{snapshot\_checksums}}
  \State \textsc{files\_modified} $\stackrel{+}\leftarrow$ \textsc{f}
  \State \textsc{bytes\_lost} $\gets$ \textsc{bytes\_lost} $+$ lookup\_size(\textsc{f})
\EndIf

\State \textsc{ptr} $\gets$ \textsc{ptr}$+1$
\EndWhile
\end{algorithmic}
\end{algorithm}

After a maximum of five minutes after the sample has entered the encryption phase, the experiment is concluded, and the snapshot of the underlay is created. To assess the damage done by the sample, Algorithm~\ref{alg:loss} computes the number of bytes lost. First, a list of checksums is computed based on the file contents in the snapshot after the experiment. Then, the same is done for the files in the initial dataset. Finally, for each file in the initial dataset, it is checked whether the checksum is contained in the post-experiment checksums. If not, the file size of the original file is assessed and added to the final result number.
\setlength{\tabcolsep}{2pt}
\begin{table*}[t]
\centering
\caption{Data Lost (in kilobytes) per Configuration and Ransomware}
\begin{tabular}{@{}ll|cccccccc|c@{}}
\toprule
            \text{Scenario} & Delay & \textit{Babuk} & \textit{Blackbasta}  & \textit{Cl0p} & conti    & \textit{GoCry} &  \textit{javaRansomware}   &    \textit{lollocker} &   \textit{Monti} & \textbf{Average}  \\ \midrule    

    \texttt{NO DEFENSE} & 0 & 10'655'744 & 10'655'744  &  10'655'744 & 561'152 &   10'655'744 & 2'634'752 & 2'926'592 & 10'655'744 & 7'425'152 \\
    \texttt{PKILL} & 0 &107'752 & 9'168 & 668'159 & 49'255&     0  & 39'762 & 620'125 & 28'889 & 190'388  \\
    \texttt{OBF} & 0 & 77'172  & 9168  & 70'644  & 96'371  &  0  & 48'500   & 137'304  &  88'259 & 65'931 \\
    \texttt{DEL+OBF} & 1 s & 1'286  & 8'266  & 19'598   & 69  & 0  & 0  & 1'692  &  69 & 3873 \\
    \texttt{DEL+OBF} &5 s & 1'289  & 4'313  &  0  & 0 &  0 & 0 & 19  & 0  & 703  \\
    \texttt{DEL+OBF} &10 s & 1'286  & 4'313  &  387  & 19  & 0  & 0  & 19  &  10 &  754  \\
    \texttt{TRACK+OBF} & <5s & 27'025  & 8'781  & 0  & 42'427 & 0  &12'149 & 29'777  & 19'489  & 17'456  \\

           \midrule
\end{tabular}
\label{table:byteslost}
\end{table*}

In that sense, the amount of data loss is computed on a pessimistic approach. For example, if only a single bit of the file is modified, the whole file is considered lost since no assumptions on the type of data are made. This also presents the danger that some malware samples may appear stronger than they realistically are. For example, some samples could delete many files without encrypting them, which would be computationally cheap. Nevertheless, it is assumed that this computation of the file modifications presents a fair approximation of data loss. It is key to highlight that only \textit{data loss} is quantified -- other impacts, such as loss of confidentiality from data access, are out of scope.

As evident from the first row in \tablename~\ref{table:byteslost} where no security mechanisms are present, most samples achieve full encryption of the $\approx$ 10 GB of data in the system. Still, there are differences in terms of data loss since certain samples focus on a dedicated set of file types. On average, 7.43 GB of data is encrypted or lost when the samples are not interrupted.
The second baseline measurement (\ie a defense strategy that has already been explored in research and thus implemented for comparative purposes) is the \texttt{PKILL} defense, which enacts process termination upon detection. The first observation for this defense strategy is that even without a novel defense mechanism, reactive detection can lead to the large majority of data being protected, as 97\% of data remains unmodified compared to the uninterrupted case.
Nevertheless, some malware samples can still destroy multiple hundred Megabytes of data until the process is positively classified and terminated. For example, \textit{Blackbasta} encrypts roughly 9 KB of data until mitigation. In these cases, it is likely that this defense approach can be truly autonomous, and some degree of administrative intervention (\eg decommissioning the device, restoring backups) is needed. This shows that by itself, the delay for detection should be further optimized to save more data. Furthermore, there is a clear termination signal that the ransomware could leverage for self-adaptation, motivating the need for additional measures.

Next, \texttt{OBF} presents a different mitigation approach, which also operates in a non-blocking manner (\ie the behavior until the first monitoring cycle does not face interference). Nevertheless, this defense configuration can reduce the data loss by $\approx$ 65\% compared to \texttt{PKILL}. This may indicate that the obfuscating defense is more suppressive against the malicious sample. Furthermore, looking into the modification times in the snapshot, it is revealed that after the detection of the ransomware, the ransomware continues to execute without any data loss. This indicates that the defense is indeed stealthy (\ie the attacker does not receive an immediate signal that it is being mitigated), which could prevent sophisticated ransomware such as roar from improving.

To improve the damage dealt until the detection system raises the alarm and deploys the defense, the three variants of \texttt{DEL+OBF} (\ie ones blocking for 1, 5, and 10 seconds) all show another strong improvement compared to either killing the process or just obfuscating file system responses. First, based on a 1-second timer, only 3.87 MB of data is lost on average for the whole experiment -- $\approx$ 98\% less than in the \texttt{PKILL} defense. The main reason this residual data is lost is that the initial delay is not long enough for the detection system. This also explains why the 5 and 10-second timers can save an additional $\approx$ 81.85\% of data, comparing the average data loss to the 1-second timer. Naturally, some data is lost, as the detection system does not perform perfectly for all samples. Furthermore, the encryption windows may not be perfectly aligned with the monitoring windows (\eg the first 5-second window may contain the first 100 ms of encryption towards the end). Interestingly, using a larger time window does not improve the defense.

Although the strategies that incorporate both obfuscation after an initial delay clearly present the most robust defense guarantees, they do so at the cost of an increased delay. Thus, application scenarios where persistence delay is critical would suffer from this strategy. To present a hybrid solution, \texttt{TRACK+OBF} applies only an initial delay for new processes. Thus, delayed critical applications would only suffer from a single performance hit, and subsequent operations can be services like benign applications. Of course, this comes at the cost that the defense is weakened. If ransomware first exhibits a benign behavior and then turns to encryption, this behavior is only detected with the delay of the monitoring and detection cycle. As shown in the last row, this is the case, as $\approx$ 17.46 MB of data are lost on average. Thus, it outperforms the remaining non-blocking defense mechanisms while underperforming against the ones delaying the execution for improved detection. Furthermore, it does so at the cost of increased complexity since the state must be maintained longer than the monitoring cycle.

In summary, the proposed defense methods provide an improved defense system. For workloads that require high-security guarantees while being able to sacrifice delay requirements, the \texttt{DEL+OBF} defense for a 5-second timer is the best choice. If the application should still perform with low delay, the \texttt{TRACK+OBF} and \texttt{OBF} strategies could be considered, as indicated in Table~\ref{table:byteslost}. The former presents stronger security guarantees but at increased complexity for managing the state and thus increased resource requirements.

 \begin{figure*}[pos=b]
     \begin{subfigure}[b]{0.329\textwidth}
          \centering
          \resizebox{\linewidth}{!}{\begin{tikzpicture}
\begin{axis}[
    xlabel={Time [s]},
    ylabel={Resource Usage [\%]},
    xmin=0, xmax=350,
    ymin=-5, ymax=100,
    mark repeat={3},
    legend style={nodes={scale=0.8, transform shape}}, 
    xtick={0, 100, 200, 300},
    ytick={0, 25, 50, 75, 100},
    legend pos=north east,
    width=\linewidth,
    legend cell align={left},
    height=4.6cm,
    ymajorgrids=true,
    grid style=dashed,
]

\addplot[
    ]
    coordinates {
    (0,0.0)(10,0.0)(21,0.0)(31,0.0)(41,0.0)(51,0.0)(61,0.0)(72,0.0)(82,0.0)(92,0.0)(102,0.0)(112,0.0)(123,0.0)(133,6.7)(143,0.0)(153,0.0)(164,0.0)(174,0.0)(184,0.0)(194,0.0)(204,0.0)(215,6.7)(225,0.0)(235,0.0)(245,0.0)(255,0.0)(266,0.0)(276,0.0)(286,0.0)(296,0.0)(307,0.0)(317,0.0)(327,0.0)(337,0.0)(347,0.0)(358,0.0)(368,0.0)
    };
    \addlegendentry{CPU (F)}

\addplot[
    legend style={at={(0.02,0.02)},anchor=south west},
    color=black,
    mark=square,
    ]
    coordinates {
(0,0.3)(10,0.3)(21,0.3)(31,0.3)(41,0.2)(51,0.4)(61,0.4)(72,0.5)(82,0.6)(92,0.7)(102,0.7)(112,0.7)(123,0.7)(133,0.7)(143,0.8)(153,0.8)(164,0.8)(174,0.8)(184,0.8)(194,0.8)(204,0.8)(215,0.8)(225,0.8)(235,0.8)(245,0.8)(255,0.8)(266,0.8)(276,0.8)(286,0.8)(296,0.8)(307,0.8)(317,0.8)(327,0.8)(337,0.8)(347,0.8)(358,0.8)(368,0.8)
    };
    \addlegendentry{RAM (F)}
\addplot[
    legend style={at={(0.02,0.02)},anchor=south west},
    color=black,
    dashed,
    mark=star,
    ]
    coordinates {
    (51,0.0)(61,0.0)(72,0.0)(82,0.0)(92,0.0)(102,0.0)(112,0.0)(123,0.0)(133,0.0)(143,0.0)(153,0.0)(164,0.0)(174,0.0)(184,0.0)(194,0.0)(204,0.0)(215,0.0)(225,0.0)(235,0.0)(245,0.0)(255,0.0)(266,0.0)(276,0.0)(286,0.0)(296,0.0)(307,0.0)(317,0.0)(327,0.0)(337,0.0)(347,0.0)
    };
    \addlegendentry{CPU (R)}
    
\addplot[
    legend style={at={(0.02,0.02)},anchor=south west},
    color=black,
    mark=triangle,
    ]
    coordinates {
    (51,7.9)(61,7.8)(72,7.8)(82,7.8)(92,7.8)(102,7.8)(112,7.8)(123,7.8)(133,7.8)(143,7.8)(153,7.8)(164,7.8)(174,7.8)(184,7.8)(194,7.8)(204,7.8)(215,7.8)(225,7.8)(235,7.8)(245,7.8)(255,7.8)(266,7.8)(276,7.8)(286,7.8)(296,7.8)(307,7.8)(317,7.8)(327,7.8)(337,7.8)(347,7.8)
    };
    \addlegendentry{RAM (R)}
\end{axis}
\end{tikzpicture}}  
          \caption{\texttt{DEL+OBF [1s]} Defense}
          \label{fig:perf_DEL_OBF_1}
     \end{subfigure}
     \begin{subfigure}[b]{0.329\textwidth}
          \centering
          \resizebox{\linewidth}{!}{\begin{tikzpicture}
\begin{axis}[
    xlabel={Time [s]},
    ylabel={Resource Usage [\%]},
    xmin=0, xmax=350,
    ymin=-5, ymax=100,
    mark repeat={3},
    xtick={0, 100, 200, 300},
    ytick={0, 25, 50, 75, 100},
    legend pos=north east,
    width=\linewidth,
    legend cell align={left},
    height=4.6cm,
    ymajorgrids=true,
    grid style=dashed,
]

\addplot[
    ]
    coordinates {
    (10,0.0)(20,0.0)(30,0.0)(41,0.0)(51,0.0)(61,0.0)(71,0.0)(81,0.0)(92,0.0)(102,0.0)(112,0.0)(122,0.0)(133,0.0)(143,0.0)(153,0.0)(163,0.0)(173,0.0)(184,0.0)(194,0.0)(204,0.0)(214,0.0)(224,0.0)(235,0.0)(245,0.0)(255,0.0)(265,0.0)(275,0.0)(286,0.0)(296,0.0)(306,0.0)(316,0.0)(327,0.0)(337,0.0)(347,0.0)(357,0.0)(367,0.0)(378,0.0)(388,0.0)

    };
    \addlegendentry{CPU (F)}
    
\addplot[
    legend style={at={(0.02,0.02)},anchor=south west},
    color=black,
    dashed,
    mark=star,
    ]
    coordinates {
    (92,0.0)(102,0.0)(112,0.0)(122,0.0)(133,0.0)(143,0.0)(153,0.0)(163,0.0)(173,0.0)(184,0.0)(194,0.0)(204,0.0)(214,0.0)(224,0.0)(235,6.7)(245,0.0)(255,0.0)(265,0.0)(275,0.0)(286,0.0)(296,0.0)(306,0.0)(316,0.0)(327,0.0)(337,0.0)(347,0.0)(357,0.0)(367,0.0)(378,0.0)

    };
    \addlegendentry{CPU (R)}
    
\addplot[
    legend style={at={(0.02,0.02)},anchor=south west},
    color=black,
    mark=square,
    ]
    coordinates {
    (10,0.3)(20,0.3)(30,0.3)(41,0.3)(51,0.3)(61,0.3)(71,0.3)(81,0.2)(92,0.4)(102,0.4)(112,0.4)(122,0.4)(133,0.4)(143,0.4)(153,0.4)(163,0.4)(173,0.5)(184,0.5)(194,0.5)(204,0.5)(214,0.5)(224,0.5)(235,0.6)(245,0.6)(255,0.3)(265,0.3)(275,0.4)(286,0.4)(296,0.4)(306,0.4)(316,0.4)(327,0.4)(337,0.5)(347,0.5)(357,0.5)(367,0.5)(378,0.5)(388,0.5)

    };
    \addlegendentry{RAM (F)}
    
\addplot[
    legend style={at={(0.02,0.02)},anchor=south west},
    color=black,
    mark=triangle,
    ]
    coordinates {
    (92,7.7)(102,7.7)(112,7.7)(122,7.7)(133,7.7)(143,7.7)(153,7.7)(163,7.7)(173,7.7)(184,7.7)(194,7.7)(204,7.7)(214,7.7)(224,7.7)(235,7.7)(245,7.7)(255,7.7)(265,7.7)(275,7.7)(286,7.7)(296,7.7)(306,7.7)(316,7.7)(327,7.7)(337,7.7)(347,7.7)(357,7.7)(367,7.7)(378,7.7)
    };
    \addlegendentry{RAM (R)}
    \legend{};
\end{axis}
\end{tikzpicture}}  
          \caption{\texttt{DEL+OBF [5s]} Defense}
          \label{fig:perf_DEL_OBF_5}
     \end{subfigure}
     \begin{subfigure}[b]{0.329\textwidth}
          \centering
          \resizebox{\linewidth}{!}{\begin{tikzpicture}
\begin{axis}[
    xlabel={Time [s]},
    ylabel={Resource Usage [\%]},
    xmin=0, xmax=350,
    ymin=-5, ymax=100,
    mark repeat={3},
    xtick={0, 100, 200, 300},
    ytick={0, 25, 50, 75, 100},
    legend pos=north east,
    width=\linewidth,
    legend cell align={left},
    height=4.6cm,
    ymajorgrids=true,
    grid style=dashed,
]

\addplot[
    ]
    coordinates {
    (10,0.0)(21,0.0)(31,0.0)(41,0.0)(51,0.0)(61,0.0)(72,0.0)(82,0.0)(92,0.0)(102,0.0)(112,0.0)(123,0.0)(133,0.0)(143,0.0)(153,0.0)(163,0.0)(174,0.0)(184,0.0)(194,0.0)(204,0.0)(215,0.0)(225,0.0)(235,0.0)(245,0.0)(255,0.0)(266,0.0)(276,0.0)(286,0.0)(296,0.0)(307,0.0)(317,0.0)(327,0.0)(337,0.0)(347,0.0)(358,0.0)
    };
    \addlegendentry{CPU (F)}
    
\addplot[
    legend style={at={(0.02,0.02)},anchor=south west},
    color=black,
    dashed,
    mark=star,
    ]
    coordinates {
    (41,0.0)(51,0.0)(61,0.0)(72,0.0)(82,0.0)(92,0.0)(102,0.0)(112,0.0)(123,0.0)(133,0.0)(143,0.0)(153,0.0)(163,0.0)(174,0.0)(184,0.0)(194,0.0)(204,0.0)(215,0.0)(225,0.0)(235,0.0)(245,0.0)(255,0.0)(266,0.0)(276,0.0)(286,0.0)(296,0.0)(307,0.0)(317,0.0)(327,0.0)(337,0.0)
    };
    \addlegendentry{CPU (R)}
    
\addplot[
    legend style={at={(0.02,0.02)},anchor=south west},
    color=black,
    mark=square,
    ]
    coordinates {
(10,0.3)(21,0.2)(31,0.1)(41,0.3)(51,0.3)(61,0.3)(72,0.3)(82,0.3)(92,0.3)(102,0.3)(112,0.3)(123,0.3)(133,0.3)(143,0.3)(153,0.3)(163,0.3)(174,0.3)(184,0.4)(194,0.4)(204,0.4)(215,0.4)(225,0.4)(235,0.4)(245,0.4)(255,0.4)(266,0.4)(276,0.4)(286,0.4)(296,0.4)(307,0.4)(317,0.4)(327,0.4)(337,0.5)(347,0.5)(358,0.5)
    };
    \addlegendentry{RAM (F)}
    
\addplot[
    legend style={at={(0.02,0.02)},anchor=south west},
    color=black,
    mark=triangle,
    ]
    coordinates {
    (41,8.5)(51,8.5)(61,8.5)(72,8.5)(82,8.5)(92,8.5)(102,8.5)(112,8.5)(123,8.5)(133,8.5)(143,8.5)(153,8.5)(163,8.5)(174,8.5)(184,8.5)(194,8.5)(204,8.5)(215,8.5)(225,8.5)(235,8.5)(245,8.5)(255,8.5)(266,8.5)(276,8.5)(286,8.5)(296,8.5)(307,8.5)(317,8.5)(327,8.5)(337,0.0)
    };
    \addlegendentry{RAM (R)}
    \legend{};
\end{axis}
\end{tikzpicture}} 
          \caption{\texttt{DEL+OBF [10s]} Defense}
          \label{fig:perf_DEL_OBF_10}
     \end{subfigure}
          \begin{subfigure}[b]{0.329\textwidth}
          \centering
          \resizebox{\linewidth}{!}{\begin{tikzpicture}
\begin{axis}[
    xlabel={Time [s]},
    ylabel={Resource Usage [\%]},
    xmin=0, xmax=150,
    ymin=-5, ymax=100,
    xtick={0, 50, 100, 150, 200},
    ytick={0, 25, 50, 75, 100},
    legend pos=north east,
    legend style={nodes={scale=0.8, transform shape}}, 
    width=\linewidth,
    legend cell align={left},
    height=4.6cm,
    ymajorgrids=true,
    grid style=dashed,
]

\addplot[
    ]
    coordinates {
    (8,0.0)(18,0.0)(29,0.0)(39,0.0)(49,0.0)(59,0.0)(69,46.7)(80,0.0)(90,0.0)(100,0.0)(110,0.0)(120,0.0)(130,0.0)(140,0.0)
    };
    \addlegendentry{CPU (F)}
    
\addplot[
    legend style={at={(0.02,0.02)},anchor=south west},
    color=black,
    dashed,
    mark=star,
    ]
    coordinates {
    (69,20.0)
    };
    \addlegendentry{CPU (R)}
    
\addplot[
    legend style={at={(0.02,0.02)},anchor=south west},
    color=black,
    mark=square,
    ]
    coordinates {
    (8,0.3)(18,0.3)(29,0.3)(39,0.3)(49,0.3)(59,0.3)(69,0.8)(80,0.9)(90,0.9)(100,0.8)(110, 0.9)(120, 0.9)(130, 0.9)(140, 0.9)
    };
    \addlegendentry{RAM (F)}
    
\addplot[
    legend style={at={(0.02,0.02)},anchor=south west},
    color=black,
    mark=triangle,
    ]
    coordinates {
    (69,8.4)
    };
    \addlegendentry{RAM (R)}

    \legend{};
    
\end{axis}
\end{tikzpicture}}  
          \caption{\texttt{PKILL} Defense}
          \label{fig:perf_PKILL}
     \end{subfigure}
     \begin{subfigure}[b]{0.329\textwidth}
          \centering
          \resizebox{\linewidth}{!}{\begin{tikzpicture}
\begin{axis}[
    xlabel={Time [s]},
    xmin=0, xmax=150,
    ylabel={Resource Usage [\%]},
    ymin=-5, ymax=100,
    xtick={0, 50, 100, 150, 200},
    ytick={0, 25, 50, 75, 100},
    legend pos=north east,
    width=\linewidth,
    legend cell align={left},
    height=4.6cm,
    ymajorgrids=true,
    grid style=dashed,
]

\addplot[
    ]
    coordinates {
    (0,0.0)(10,0.0)(20,0.0)(30,0.0)(41,0.0)(51,73.3)(61,73.3)(71,60.0)(82,66.7)(92,73.3)(102,73.3)(112,0.0)(122,0.0)(133,0.0)
    };
    \addlegendentry{CPU (F)}
    
\addplot[
    legend style={at={(0.02,0.02)},anchor=south west},
    color=black,
    dashed,
    mark=star,
    ]
    coordinates {
    (51,40.0)(61,33.3)(71,33.3)(82,26.7)(92,33.3)(102,33.3)
    };
    \addlegendentry{CPU (R)}
    
\addplot[
    legend style={at={(0.02,0.02)},anchor=south west},
    color=black,
    mark=square,
    ]
    coordinates {
    (0,0.3)(10,0.3)(20,0.2)(30,0.2)(41,0.1)(51,0.9)(61,0.9)(71,0.9)(82,0.8)(92,0.8)(102,0.8)(112,0.8)(122,0.8)(133,0.8)
    };
    \addlegendentry{RAM (F)}
    
\addplot[
    legend style={at={(0.02,0.02)},anchor=south west},
    color=black,
    mark=triangle,
    ]
    coordinates {
    (51,7.9)(61,7.8)(71,7.8)(82,7.9)(92,7.6)(102,7.6)
    };
    \addlegendentry{RAM (R)}
    \legend{};

\end{axis}
\end{tikzpicture}}  
          \caption{\texttt{OBF} Defense}
          \label{fig:perf_OBF}
     \end{subfigure}
     \begin{subfigure}[b]{0.329\textwidth}
          \centering
          \resizebox{\linewidth}{!}{\begin{tikzpicture}
\begin{axis}[
    xlabel={Time [s]},
    ylabel={Resource Usage [\%]},
    xmin=0, xmax=150,
    ymin=-5, ymax=100,
    xtick={0, 50, 100, 150, 200},
    ytick={0, 25, 50, 75, 100},
    legend pos=north east,
    width=\linewidth,
    legend cell align={left},
    height=4.6cm,
    ymajorgrids=true,
    grid style=dashed,
]

\addplot[
    ]
    coordinates {
(0,0.0)(10,0.0)(20,0.0)(30,60.0)(40,73.3)(51,66.7)(61,73.3)(71,66.7)(81,66.7)(92,0.0)(102,0.0)(112,0.0)(122,0.0)(132,0.0)
    };
    \addlegendentry{CPU (F)}
    
\addplot[
    legend style={at={(0.02,0.02)},anchor=south west},
    color=black,
    dashed,
    mark=star,
    ]
    coordinates {
(20,46.7)(30,26.7)(40,33.3)(51,26.7)(61,33.3)(71,26.7)(81,33.3)
    };
    \addlegendentry{CPU (R)}
    
\addplot[
    legend style={at={(0.02,0.02)},anchor=south west},
    color=black,
    mark=square,
    ]
    coordinates {
(0,0.3)(10,0.3)(20,0.2)(30,0.9)(40,0.9)(51,0.9)(61,0.9)(71,0.8)(81,0.9)(92,0.9)(102,0.9)(112,0.9)(122,0.9)(132,0.9)
    };
    \addlegendentry{RAM (F)}
    
\addplot[
    legend style={at={(0.02,0.02)},anchor=south west},
    color=black,
    mark=triangle,
    ]
    coordinates {
(20,3.7)(30,8.9)(40,8.8)(51,8.8)(61,8.9)(71,8.6)(81,8.6)
    };
    \addlegendentry{RAM (R)}

    \legend{};
    
\end{axis}
\end{tikzpicture}}  
          \caption{\texttt{TRACK+OBF} Defense}
          \label{fig:perf_TRACK_OBF}
     \end{subfigure}
    \caption{Performance Analysis of Different Defense Methods and Configurations against \textit{javaRansomware}}
    \label{fig:resource-consumed}
 \end{figure*}
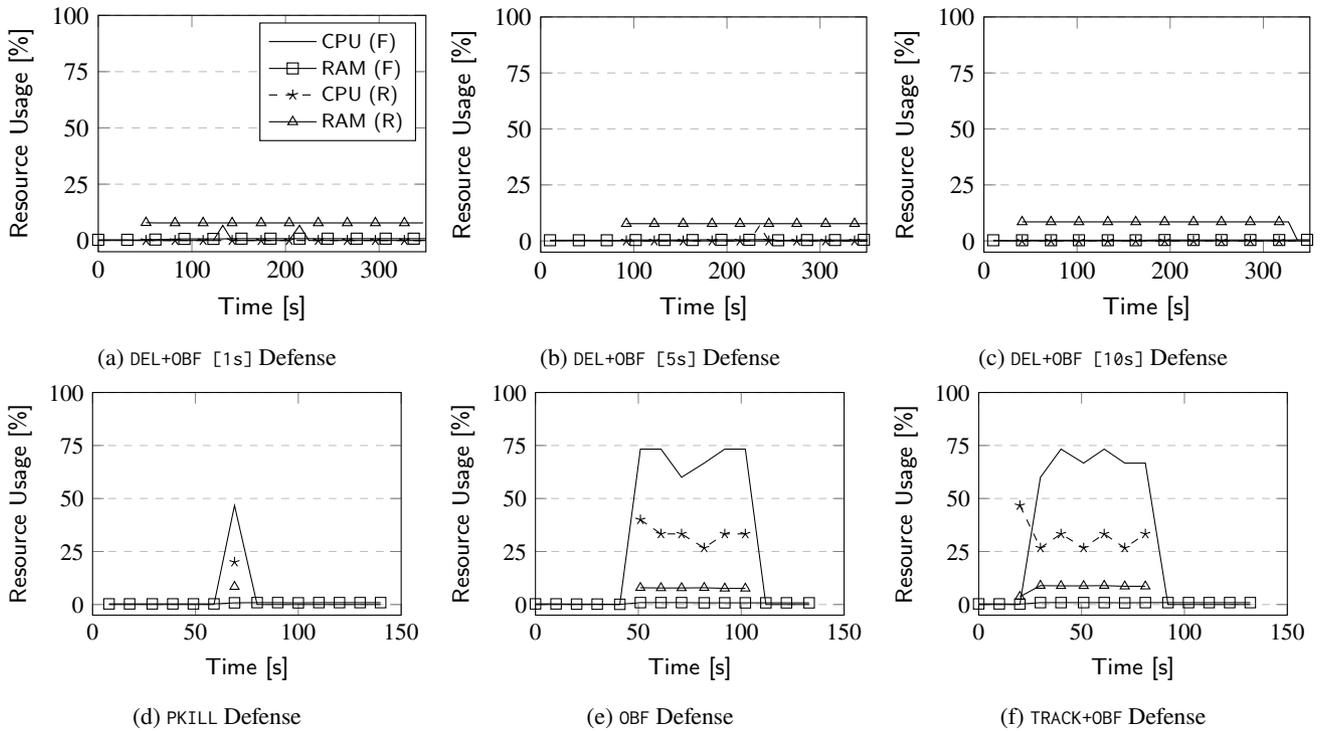
 

\subsubsection{Resource Efficiency}
Although all defense mechanisms present a certain improvement in resilience against ransomware attacks, they lead to different behaviors from the offensive (\ie blocking or non-blocking encryption) and defensive (\ie stateless or stateful) perspective. In \figurename~\ref{fig:resource-consumed}, the analysis of executing the \textit{javaRansomware} sample against the different defense configurations regarding resource consumption is visualized. For each experiment, the relative CPU consumption and the reserved system memory are visualized. These two metrics are assessed for the process (and subprocesses) spawned by the ransomware and for the file system, which includes the defense and detection planes running in the background.

The first observation is that, due to the delaying aspect of the \texttt{DEL+OBF} defense family, the overall resource consumption is smoothed, resulting in less bursty resource consumption. However, it is important not to confuse "less bursty" with "fewer resources consumed" -- overall, the same resources are consumed since the same workload is performed. Still, the ransomware iterates through all the buffers representing the protected files, encrypts them, and attempts to write them to disk. In that sense, delaying and obfuscating the modifying operations (\ie rename, delete, write), not only mitigates the modifications but also slows down the ransomware.

Since the \texttt{OBF} and \texttt{TRACK+OBF} defenses do not extensively delay the ransomware execution but instead try to mimic the file system in its normal state, there is substantially bursty resource consumption. Here, it can be seen that keeping up with computationally intensive ransomware leads to significant resource consumption by the file system, too. However, assuming that ransomware attacks are still rare, it could be argued that this resource consumption can be afforded. Interestingly, the execution of the ransomware does not change once the file system mitigation goes into action, emphasizing that the ransomware may not detect the mitigation. This is magnified by the fact that the ransomware stops the encryption process by itself once the target data is consumed, encrypted, and disk persistence is attempted.

If the \texttt{DEL+OBF} strategies are compared to the \texttt{PKILL} defense, it is clear that these novel defense mechanisms are not more resource-effective. Indeed, \texttt{PKILL} is the most resource-effective approach since only a few monitoring cycles exhibit ransomware behavior. However, this comes at a cost, where it is assumed that \1 the ransomware does not alter its behavior based on the \texttt{KILL} signal and \2 that it is actually possible to prevent further execution of the sample. Here, the defense methods introduced in this paper present the advantage of being stealthy while avoiding the damage function of the ransomware, even if the sample may not be removable.

In summary, the \texttt{PKILL} defense is the most resource-effective one, although at the expense of weak security (\ie assuming killing the process mitigates the malware). The other strategies require roughly the same amount of resources, although the ones involving a delay component smooth the consumption over time while providing a slightly stronger defense.

\subsubsection{Usability in Benign Workloads}
Although efficient resource usage when mitigating ransomware is important, efficient operation in benign settings is also critical since devices likely spend much more time in a non-infected state. Thus, efficiency optimizations when running benign workloads are an important pillar to ensure the overall efficiency of the solution. Previous experiments demonstrated that the \texttt{TRACK+OBF} defense presents strong defense effectiveness. Due to its design optimizations concerning efficiency, this defense mechanism is deployed in this scenario while confronting it against several additional workloads. The following workloads are then measured concerning their execution time and the resources required by the workload and the defense system. For each workload, these values are established when running them in the overlay system without any monitoring, detection, or mitigation system running and when deploying them with the aforementioned defense strategy (see Table~\ref{table:benign}):

\begin{itemize}
    \item In the first workload \texttt{WL1}, the effect of the system on system administration is considered. Thus, the packages for \textit{apache2}, \textit{BIND9}, and \textit{MariaDB} servers are downloaded and subsequently installed into the overlay file system. This task involves many (concurrent) read and write operations involving high entropy data.
    \item For \texttt{WL2}, a reading from a real air quality sensor is retrieved and stored on the disk. Thus, this workload shows low read operations, a single longrunning write operation with low delay requirements.
    \item The third workload \texttt{WL3}, involves the creation of a backup -- here, an archive of 977 files, protected by the file system, is created. The resulting archive is stored as a GZIP-compressed tarball on the same file system.
    \item In \texttt{WL4}, a long-running, write-intensive task is carried out, which involves continuous download of the \textit{libreoffice} suite into the file system. Here, it is also investigated how the system performs when there are no other bottlenecks, leading to highly asynchronous operation. Thus, the data is downloaded over a local network link provisioned at 1 Gbit/s. In that sense, writing network data at such a high throughput also resembles a stress test for the file system.
\end{itemize}

\begin{table}[pos=b]
\centering
\caption{Overhead on Benign Workloads}
\label{table:benign}
\begin{tabular}{@{}lrllll@{}}
\toprule    
            \textit{}  
            & \textit{} &      \multicolumn{2}{c}{\textit{CPU} [\%]} &  \multicolumn{2}{c}{\textit{RAM} [MB]} \\
            
             \textit{Scenario}& \textit{Time} [s] &  \textit{System} & \textit{GuardFS} & \textit{System} & \textit{GuardFS}
            \\ \midrule
            WL1 & 3  & 19.7 & - & 2.6 & -\\
            WL1+\texttt{TRACK+OBF} & 8  & 13.4 & 56.5 &  1.8 & 10.8\\
             \midrule
            WL2 & 4 & 1 & - & 4.8 & -\\
            WL2+\texttt{TRACK+OBF} & 6 & 1 & 1 & 4.9 & 14.1 \\
             \midrule
            WL3 & 16 & 51.1 & - & 13.43 & -\\
            WL3+\texttt{TRACK+OBF} & 20 & 33.33 & 1 & 1.9 & 12.1 \\
             \midrule
            WL4 & 23 & 9.8 & - & 10.5 & -\\
            WL4+\texttt{TRACK+OBF} & 24 & 2.24 & 50.1 & 2.25 & 7.28  \\
             \bottomrule
            
\end{tabular}
\end{table}

Looking at the \textit{Time} column, it can be seen that the defense and monitoring system does lead to a delay in execution time. This comes as expected, as this defense configuration involves an initial delay for each process (and any spawned child processes) to classify the first interaction with the file system. This efficiency is least optimal for short-running processes that involve the creation of many files by numerous parallel processes, as represented by \texttt{WL1} in the first row. This effect becomes less prominent for workloads like \texttt{WL4} that are long-running and involve continuous write operations. Furthermore, for workloads such as \texttt{WL2}, that are not delay-critical and operating in a highly asynchronous manner (\ie spending a lot of time waiting for other I/O tasks such as network requests), this constraint may likely not play a difference. In such cases, the added defense effectiveness could provide a credible tradeoff between usability and security. Another observation is that even for creating high entropy data, such as in \texttt{WL3}, no false alerts (and thus mitigation) were raised during regular operation. Regarding resource consumption, it can be seen that for short-running, bursty workloads involving many system calls, the file system requires substantial compute resources and static memory consumption in the magnitude of roughly 10 MB.

\subsection{Comparison with Related Work}
In the proactive approach proposed in~\cite{mtfs}, a file-system-based ransomware mitigation is described, enabling a comparison of reactive and proactive paradigms in autonomous ransomware defense. As highlighted by the experiments that assess the overhead in benign settings, the reactive approach shown in this paper does present one issue. To be able to react to attacks, data must be monitored continuously. In this paper, file system-related system call parameters were considered. This makes detection robust. However, processing system calls becomes more expensive as more system calls are created. In that sense, the higher the load on a system, the more resources are required to assess the system calls. This monitoring cost is avoided in a proactive defense since the defense techniques are deployed beforehand.

To make proactive defense viable, deploying such a defense strategy imposes constraints on how the defense can be designed. After all, the defense strategy is constantly executing, which leads to overhead created by the mitigation. In~\cite{icc}, the proactive defense was found too expensive, especially for ransomware, since deception was achieved by creating a set of actual files to trap the ransomware encryption. Thus, only a lightweight defense mechanism was considered suitable for proactive deployment. In that sense, \cite{mtfs}~presents a lightweight defense mechanism that can be deployed without intelligence. However, implementing it relies on specific assumptions of adversarial behavior, thus weakening the defense's effectiveness. For example, one of the techniques relies on the ransomware performing a full traversal of the target file system before attempting any encryption. Although this was demonstrated to be effective, it could potentially be circumvented by changing the traversal strategy or by performing traversal in parallel to the encryption. Indeed, the defense mechanisms presented here also rely on the attacker performing encryption (and thus creating a specific pattern of read and write operation, as well as creating high entropy data as output). However, as discussed in the history of ransomware, crypto-ransomware is still a highly relevant threat vector and has been so for a long time. This is backed by the results on the defense effectiveness, which show that the proactive approach led to $\approx$ 300 MB being lost, while the experiments on this reactive deployment showed losses between $\approx$ 66 MB for the worst case and $\approx$ 0.7 MB for the best strategy. In summary, reactive mitigation can provide optimized defense, although resources for continuous monitoring, processing, and classification are needed. Concerning the framework architecture presented, the latter two steps could be centralized in an external device.

\section{Conclusion}
\label{sec:conclusion}
This work presented the design and prototypical implementation of an integrated detection and defense platform that leverages the file system to provide autonomous and fully automated mitigation against ransomware. The detection plane relies on system call data that can be intercepted on the file system level of the operating system. With this data, an ML-based binary classifier can deploy different strategies in the defense plane. Then, multiple novel mechanisms involving stealthy defense have been proposed. Finally, a set of experiments has shown the performance of the detection plane in offline and online tests, while the detection and mitigation capabilities were assessed against several ransomware samples in a real scenario and using a virtualized testbed. Here, the defense effectiveness, resource consumption, and side effects on benign workloads were studied, leading to a comparison of proactive defense solutions with a data-driven, reactive defense.

In conclusion, this work demonstrated that ML-based reactivity can optimize the defense capabilities of a defense system. Depending on the security requirements, the defense strategy with the highest robustness could provide an almost completely automated defense system, with no manual intervention required after successful detection. It is critical to select the appropriate defense configuration depending on the type of workload to be performed in the benign setting. For highly delay-critical, and short-lived processes, the most complex defense method \texttt{TRACK+OBF} is suitable. In contrast, ones that can sacrifice delay for improved security benefit from the \texttt{DEL+OBF} mechanisms. These two also show different effects on resource consumption, with the second one being less bursty. Finally, from the detection plane, it can be concluded that the ML-based classifier presents robust detection for a myriad of different ransomware samples, even when behavioral data was gathered for different (and thus unseen) samples. Still, monitoring system-call data comes at a cost, especially at high load, where the number of system calls to be processed increases.

Based on the experiences drawn, multiple avenues for further research are identified. First, it will be investigated how the detection plane could be made more lightweight for scenarios where data processing cannot be offloaded. Here, different data sources, such as performance metrics will be analyzed. Furthermore, the platform will be tested using other benign workloads (\eg office usage, IoT scenarios) and additional ransomware samples. Here, the portability of the platform to other operating systems will be investigated.

\section*{Declaration of Competing Interest}
The authors declare that they have no known competing financial interests or personal relationships that could have appeared to influence the work reported in this paper. 

\section*{CRediT authorship contribution statement}

\textbf{Jan von der Assen.} Methodology, Conceptualization, Writing - Review \& Editing.
\textbf{Chao Feng.} Writing, Review. 
\textbf{Alberto Huertas Celdran.} Methodology, Writing - Review. 
\textbf{Robert Oles.} Data curation, Analysis.
\textbf{Gérôme Bovet.} Project administration, Funding acquisition.
\textbf{Burkhard Stiller.} Supervision, Funding acquisition.

\section*{Acknowledgment}

This work has been partially supported by \textit{(a)} the Swiss Federal Office for Defense Procurement (armasuisse) with the CyberMind project (CYD-C-2020003) and \textit{(b)} the University of Zürich UZH.

%
%
\balance
\bibliographystyle{cas-model2-names}
\bibliography{main}

\bio{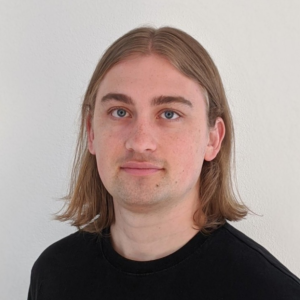} Jan von der Assen received his MSc degree in Informatics from the University of Zurich. Currently, he is pursuing his Doctoral Degree under the supervision of Prof. Dr. Burkhard Stiller at the Communication Systems Group, University of Zurich. His research interest lies at the intersection between risk management and the mitigation of cyber threats. Contact him at vonderasssen@ifi.uzh.ch \endbio

\bio{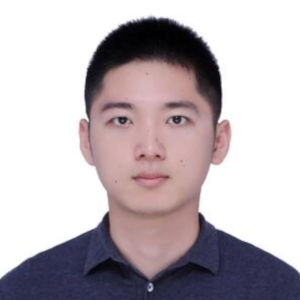} Chao Feng received the MSc degree in Informatics from the University of Zurich, Switzerland. He is currently pursuing his Ph.D. in computer science at the Communication Systems Group, Department of Informatics at the University of Zurich. His scientific interests include IoT, cybersecurity, data privacy, machine learning, and computer networks. Contact him at cfeng@ifi.uzh.ch \endbio

\bio{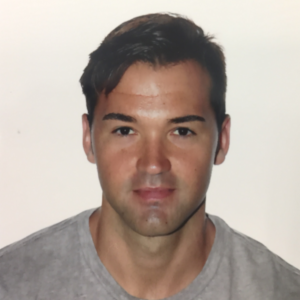} Alberto Huertas Celdrán is a senior researcher at the Communication Systems Group CSG, Department of Informatics IfI, University of Zurich UZH. He received his PhD degree in Computer Science from the University of Murcia, Spain. His scientific interests include cybersecurity, machine and deep learning, continuous authentication, and computer networks. Contact him at huertas@ifi.uzh.ch \endbio

\bio{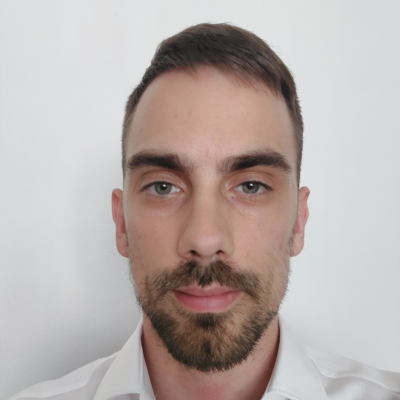} Róbert Oleš holds a Bachelor's degree in Economics and a Master's in Computer Science. Previously, he worked as a data scientist with a focus on large-scale geospatial projects. Robert currently works as a software engineer. His interests lie in distributed systems, software engineering, and predictive data analysis. Contact him at olesrobert1@gmail.com \endbio

\bio{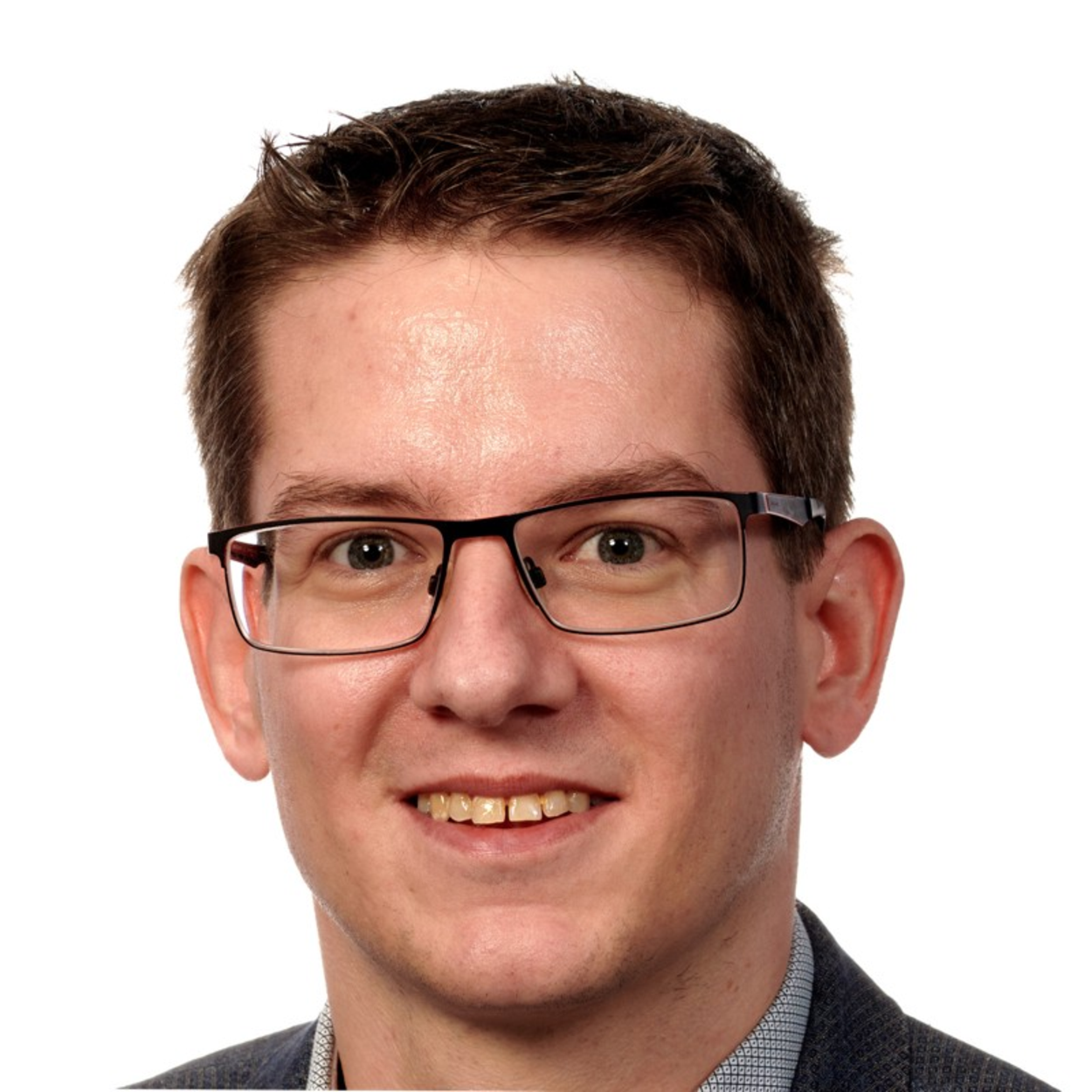} Gérôme Bovet is the head of data science for the Swiss Department of Defense. He received his PhD in networks and computer systems from Telecom ParisTech, France. His work focuses on Machine and Deep Learning, with an emphasis on anomaly detection, adversarial and collaborative learning in IoT sensors. Contact him at gerome.bovet@armasuisse.ch \endbio

\bio{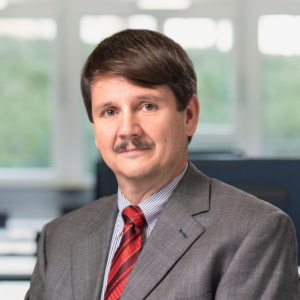} Burkhard Stiller chairs the Communication Systems Group CSG, Department of Informatics IfI, University of Zürich UZH, as a Full Professor. He received the MSc and PhD degrees from the University of Karlsruhe, Germany. His main research interests include fully decentralized systems, network and service management, IoT, and telecommunication economics. Contact him at stiller@ifi.uzh.ch \endbio

\end{document}